\documentclass[final,3p,times,twocolumn]{elsarticle}
\usepackage{epsfig}
\usepackage{amsmath}
\usepackage{color}
\usepackage{latexsym}
\usepackage{amssymb}
\usepackage{dsfont}
\usepackage{multirow}

\newcommand{\bea}{\begin{eqnarray}}
\newcommand{\eea}{\end{eqnarray}}
\newcommand{\be}{\begin{equation}}
\newcommand{\ee}{\end{equation}}

\newcommand{\ud}{\mathrm{d}}
\newcommand{\uL}{\mathcal{L}}

\newcommand{\uTr}{\mathrm{Tr}}

\newcommand{\uslash}{/\!\!\!}

\newcommand{\barpsi}{\overline{\psi}}

\newlength\savedwidth

\begin{document}
%\preprint{LBNL-xxx}\preprint{RBRC-xxx}
%\preprint{}

\title{Gauge-covariant canonical formalism revisited with application to the proton spin decomposition}

\author{C\'edric Lorc\'e}
\address{IPNO, Universit\'e Paris-Sud, CNRS/IN2P3, 91406 Orsay, France\\
        and LPT, Universit\'e Paris-Sud, CNRS, 91405 Orsay, France}

\begin{abstract}
We revisit the gauge-covariant canonical formalism by separating explicitly physical and gauge degrees of freedom. We show in particular that the gauge-invariant linear and angular momentum operators proposed by Chen \emph{et al.} can consistently be derived from the standard procedure based on the Noether's theorem. Finally, we demonstrate that this approach is essentially equivalent to the gauge-invariant canonical formalism based on the concept of Dirac variables. Because of many similarities with the background field method, the formalism developed here should also be relevant to general relativity and any metric theories.
\end{abstract}

\maketitle

\section{Introduction}

The canonical formalism allows one to derive the classical equations of motion and the conserved currents associated with the global symmetries of a Lagrangian. Unfortunately, this formalism runs into problems in presence of a gauge symmetry. In particular, the canonical energy-momentum tensor obtained following the standard procedure turns out to be gauge dependent. Since physical observables are gauge invariant, there is a widespread belief that the canonical quantities are not really physical. Moreover, the gauge dependence of the canonical variables makes the quantization procedure particularly non-trivial. Similar problems arise in any metric theories like \emph{e.g.} general relativity.

Different strategies have been adopted to deal with these issues. The oldest one goes back to Dirac \cite{Dirac:1955uv} who proposed to reformulate QED in terms of gauge-invariant fields known as \emph{Dirac variables}. These variables are constructed by adjoining phase factors to the original fields, and have been rediscovered and generalized several times under different names and in different contexts \cite{DeWitt:1962mg,Mandelstam:1962mi,Mandelstam:1968hz,Mandelstam:1968ud,BialynickiBirula:1963,Steinmann:1983ar,Steinmann:1985id,Skachkov:1985cz,Haagensen:1997pi,Horan:1998im,Masson:2010vx,Fournel:2012cr,Chen:2012vg}. We refer to \cite{Pervushin:2001kq} for a review of the subject and to \cite{Lantsman:2006ry} for a comparison with the more standard Faddeev-Popov approach. 

Schwinger \cite{Schwinger:1962zz,Schwinger:1962fg}, followed by Arnowitt and Fickler \cite{Arnowitt:1962cv}, adopted a different strategy. They proposed to separate explicitly the physical degrees of freedom from the unphysical ones in the gauge potential. The same idea has also been considered and generalized several times \emph{e.g.} in Refs.~\cite{Goto:1966,Treat:1973yc,Treat:1975dz,Duan:1979,Duan:1984cb,Duan:1998um,Duan:2002vh,Fulp:1983bt,Kashiwa:1996rs,Kashiwa:1996hp}, and reappeared more recently in the context of the gauge-invariant decomposition of the proton spin \cite{Chen:2008ag,Chen:2011zzh,Wakamatsu:2010qj,Wakamatsu:2010cb,Lorce:2012rr}. Interestingly, this approach shares many common features with the background field method introduced by DeWitt \cite{DeWitt:1967ub,DeWitt:1967uc,DeWitt:1980jv,'tHooft:1975vy,Grisaru:1975ei,Boulware:1980av,Abbott:1980hw}. The latter has been extensively used in gravity and supergravity \cite{'tHooft:1973us,'tHooft:1974bx,Deser:1977nt,Abbott:1981ff,Petrov:2007xva,Petrov:2012qn},  and in both continuum and lattice gauge theories \cite{Dashen:1980vm,Abbott:1982jh,Luscher:1995vs,Freire:2000bq,Binosi:2009qm,Binosi:2012st}. A nice introduction to the background field method can be found in \cite{Abbott:1981ke}.

To the best of our knowledge, the Schwinger approach is usually adopted either in the path-integral formalism or as an \emph{ad hoc} procedure \emph{after} the application of the standard (gauge non-invariant) canonical formalism. Surprisingly, it has never been used to develop directly a canonical formalism consistent with the gauge symmetry. In this letter, we aim at filling this gap. We show how the explicit separation of physical and gauge degrees of freedom naturally leads to a covariant form of the Euler-Lagrange equations and of the Noether's theorem. In section \ref{sec2}, we briefly remind the textbook approach to the canonical formalism. Then we present its covariant form in the presence of external or non-dynamical gauge fields, defining on the way covariant functional derivatives. In section \ref{sec3}, we develop the gauge-covariant canonical formalism based on the decomposition of the gauge field into physical and pure-gauge parts. Applying this new gauge-covariant canonical formalism to QCD, we recover the gauge-invariant decomposition of the linear and angular momentum operators constructed by Chen \emph{et al.}~\cite{Chen:2008ag,Chen:2011zzh,Wakamatsu:2010cb}. This naturally explains why their operators are the generators of translations and Lorentz transformations for the physical fields. We also comment on the lack of uniqueness of this approach and its physical relevance. In section \ref{sec4}, we discuss the Dirac's gauge-invariant canonical formalism, and demonstrate its formal equivalence with the gauge-covariant canonical formalism we propose. Finally, we conclude this letter with section \ref{sec5}.

While we confine here our discussions to the gauge theories, the same approach can easily be adapted to general relativity and any metric theories.

\section{Canonical formalism}\label{sec2}

We start with a short reminder of the standard derivation of the Euler-Lagrange equations and the Noether's theorem. Then we explain how one reconciles the standard approach with the gauge symmetry in presence of external gauge fields.

\subsection{Standard approach}\label{standardapproach}

In standard textbooks like \emph{e.g.} \cite{Ryder:1985wq}, the Lagrangian is usually thought of as a function of a generic set of fields $\phi(x)$ and their ordinary derivatives\footnote{One can also add an explicit dependence on the space-time coordinates $x$, but this does not affect in a significant way our discussions.}
\begin{equation}
\uL(x)=f[\phi(x),\partial_\mu\phi(x)].
\end{equation}
Setting to zero the variation of the action $\delta S=\int_\Omega\ud^4x\,\delta\uL(x)=0$ under arbitrary (infinitesimal) variation of the fields 
\begin{equation}
\phi(x)\mapsto\phi'(x)=\phi(x)+\delta\phi(x)
\end{equation}
that satisfy the condition $\delta\phi=0$ on the space-time boundary $\partial\Omega$, one obtains the Euler-Lagrange equations
\begin{equation}
\partial_\mu\frac{\partial\uL}{\partial(\partial_\mu\phi)}-\frac{\partial\uL}{\partial\phi}=0.
\end{equation}
In the Noether's theorem \cite{Noether:1918zz}, one considers more general variations of the fields
\begin{equation}\label{fullvar}
\begin{split}
\phi(x)\mapsto\phi'(x')&=\phi(x)+\Delta\phi(x)\\
&=\phi(x)+\delta\phi(x)+\partial_\mu\phi(x)\,\delta x^\mu,
\end{split}
\end{equation}
where $\delta\phi(x)$ represents again an intrinsic change in the functional form of the fields, $\partial_\mu\phi(x)\,\delta x^\mu$ represents a change coming from the fact that the fields are evaluated at a slightly displaced point $x'=x+\delta x$, and $\Delta\phi(x)$ is the total variation. By definition, continuous symmetries leave the action invariant\footnote{We adopt the passive point of view, so that the space-time volume is not affected.}
\begin{equation}
\delta S=\int_\Omega\ud^4x'\,\uL'(x')-\int_\Omega\ud^4x\,\uL(x)=0,
\end{equation}
where $\uL'(x')\equiv f[\phi'(x'),\partial'_\mu\phi'(x')]$. Using the Euler-Lagrange equations, one concludes that the following (infinitesimal) currents are conserved
\begin{equation}\label{current}
\mathcal J^\mu=\frac{\partial\uL}{\partial(\partial_\mu\phi)}\,\Delta\phi-\left[\frac{\partial\uL}{\partial(\partial_\mu\phi)}\,\partial_\nu\phi-\delta^\mu_\nu \uL\right]\delta x^\nu.
\end{equation}

When the Lagrangian is invariant under a gauge symmetry, the Euler-Lagrange equations turn out to be gauge covariant. Note however that the terms $\partial_\mu\frac{\partial\uL}{\partial(\partial_\mu\phi)}$ and $\frac{\partial\uL}{\partial\phi}$ are not in general separately gauge covariant. More troublesome is the fact that the currents associated with the Poincar\'{e} transformations are also not gauge invariant. It is then often claimed that the canonical linear and angular momentum operator densities have no physical significance.

\subsection{Gauge-covariant approach}

When the gauge field is treated as an \emph{external} or \emph{background} field, it is actually possible to reconcile the gauge symmetry with the Noether's theorem \cite{Ray:1968,Jackiw:1978ar,Levitsky:1981rv,Levitsky:1982mr,Hamamoto:1983fv,Barnich:1994cq}. The origin of the problem comes from the fact that the standard approach deals with quantities that are not gauge covariant, and therefore ill-defined from the geometrical point of view.

To keep the presentation simple, consider that the fields $\phi(x)$ transform as internal vectors under gauge transformations
\begin{equation}
\phi(x)\mapsto\tilde\phi(x)=U(x)\phi(x).
\end{equation}
All the following expressions can of course easily be adapted to any kinds of internal tensor transformation law. Since the original and transformed fields are evaluated at the same point, it follows that the intrinsic variation of the fields $\delta\phi(x)$ transforms also as an internal vector\footnote{Even when the transformation of the fields involves non-tensorial terms, the latter are cancelled in the intrinsic variation.} under gauge transformations. The action being gauge invariant, one can deduce from
\begin{equation}
\delta S=\int_\Omega\ud^4x\left[\frac{\partial\uL}{\partial\phi}-\partial_\mu\frac{\partial\uL}{\partial(\partial_\mu\phi)}\right]\delta\phi
\end{equation}
that the Euler-Lagrange equations are automatically gauge covariant.

In a gauge theory, the ordinary partial derivatives are not the natural geometric objects but usually appear as part of covariant derivatives $D_\mu=\partial_\mu-igA_\mu$. It is therefore more natural to consider the Lagrangian as a function of the fields and their covariant derivatives
\begin{equation}
\uL(x)=f'[\phi(x),D_\mu\phi(x)].
\end{equation}
Simple algebra shows that \cite{Lewis:2009na}
\begin{subequations}
\begin{align}
\frac{\partial f}{\partial\phi}&=\frac{\partial f'}{\partial\phi}-\frac{\partial f'}{\partial(D_\mu\phi)}\,igA_\mu,\\
\frac{\partial f}{\partial(\partial_\mu\phi)}&=\frac{\partial f'}{\partial(D_\mu\phi)}.
\end{align}
\end{subequations}
When the Lagrangian is expressed only in terms of gauge-covariant variables, the corresponding functional derivatives are automatically gauge covariant. We therefore propose to define \emph{covariant functional derivatives} and conjugate fields as follows
\begin{subequations}\label{covdef}
\begin{align}
\uL\!\stackrel{\leftarrow}{D}\!\!\!\!\!\phantom{\partial}_\phi&\equiv\frac{\partial f}{\partial\phi}+\frac{\partial f}{\partial(\partial_\mu\phi)}\,igA_\mu=\frac{\partial f'}{\partial\phi},\\
\Pi^\mu_\phi&\equiv\frac{\partial f}{\partial(\partial_\mu\phi)}=\frac{\partial f'}{\partial(D_\mu\phi)}.
\end{align}
\end{subequations}
The Euler-Lagrange equations can then be rewritten as
\begin{equation}
\frac{\partial\uL}{\partial(D_\mu\phi)}\!\stackrel{\leftarrow}{D}_\mu-\frac{\partial\uL}{\partial\phi}=0
\end{equation}
with the covariant derivative in the conjugate fundamental representation given by $\stackrel{\leftarrow}{D}_\mu=\stackrel{\leftarrow}{\partial}_\mu\!+igA_\mu$, or more compactly using the definitions \eqref{covdef}
\begin{equation}
\Pi^\mu_\phi\!\stackrel{\leftarrow}{D}_\mu\!-\uL\!\stackrel{\leftarrow}{D}_\phi=0.
\end{equation}
In this form, the individual terms of the Euler-Lagrange equations are now gauge covariant.

The problem with the Noether's theorem in the standard approach boils down to Eq. \eqref{fullvar}. When the fields transform \emph{e.g.} as internal vectors under gauge transformations, this expression does not make sense anymore from the geometrical point of view. Indeed, the fields $\phi'(x')$ and $\phi(x)$ live in different copies of the internal space, since they are evaluated at different space-time points. It then follows that, contrary to the intrinsic variation $\delta\phi(x)$, the total variation $\Delta\phi(x)$ does not transform covariantly. The consistent expression is \cite{DeWitt:2003pm,Frankel:1997ec}
\begin{equation}
\phi'(x')=\mathcal W(x+\delta x,x)\left[\phi(x)+\Delta_c\phi(x)\right],
\end{equation}
where the infinitesimal Wilson line and the covariant total variation are respectively defined as
\begin{subequations}
\begin{align}
\mathcal W(x+\delta x,x)&=1+igA_\mu(x)\,\delta x^\mu,\\
\Delta_c\phi(x)&=\delta\phi(x)+D_\mu\phi(x)\,\delta x^\mu.
\end{align}
\end{subequations}
From the gauge transformation of the $A_\mu(x)$ field
\begin{equation}\label{Agauge}
A_\mu(x)\mapsto\tilde A_\mu(x)=U(x)\left[A_\mu(x)+\frac{i}{g}\,\partial_\mu\right]U^{-1}(x),
\end{equation}
it is easy to see that the infinitesimal Wilson line transforms in a simple way
\begin{align}
\mathcal W(x+\delta x,x)\mapsto&\tilde{\mathcal W}(x+\delta x,x)\nonumber\\
&\hspace{-1cm}=U(x+\delta x)\mathcal W(x+\delta x,x)U^{-1}(x),
\end{align}
and allows one to parallel transport $\phi'(x')$  to $\phi'_\parallel(x)=\phi(x)+\Delta_c\phi(x)$ that lives in the same copy of the internal space as $\phi(x)$. The consistent expression for the conserved current is then
\begin{equation}\label{covcurrent}
\mathcal J^\mu=\frac{\partial\uL}{\partial(D_\mu\phi)}\,\Delta_c\phi-\left[\frac{\partial\uL}{\partial(D_\mu\phi)}\,D_\nu\phi-\delta^\mu_\nu \uL\right]\delta x^\nu.
\end{equation}

We stress that this approach is fine as long as the gauge field is external. But once $A_\mu(x)$ is treated as a dynamical field, the gauge-covariant formalism presented in this section cannot be applied anymore, simply because the gauge field does not transform as an internal tensor under gauge transformations. To the best of our knowledge, no consistent gauge-covariant canonical formalism with dynamical gauge field has been developed so far. We fill this gap in the next section.

\section{Gauge-covariant canonical formalism}\label{sec3}

We propose in this section, for the first time, a canonical formalism that is consistent with the gauge symmetry. Note that the gauge field plays essentially two roles. On the one hand, it is used to form a gauge-covariant derivative and to render the Lagrangian gauge invariant. On the other hand, it gives rise to a non-vanishing field strength $F_{\mu\nu}=\partial_\mu A_\nu-\partial_\nu A_\mu-ig[A_\mu,A_\nu]$ and provides the coupling with the source fields. The first aspect is somewhat unphysical as it concerns only the gauge symmetry which is not observable. On the contrary, the second aspect is physical as the field strength (or curvature) affects the trajectories of particles, and is therefore observable. We adopt here the Schwinger's strategy of separating these two aspects explicitly. 

\subsection{Decomposition of the gauge field}

Using the notation introduced by Chen \emph{et al.}, we decompose the gauge field as follows \cite{Chen:2008ag,Chen:2011zzh,Wakamatsu:2010cb}
\begin{equation}\label{Adecomp}
A_\mu(x)=A^\text{pure}_\mu(x)+A^\text{phys}_\mu(x),
\end{equation}
where $A^\text{pure}_\mu(x)$  and $A^\text{phys}_\mu(x)$ contain only gauge and physical degrees of freedom, respectively. By definition, the pure-gauge field is unphysical and therefore cannot contribute to the field strength
\begin{equation}\label{Apuredef}
F^\text{pure}_{\mu\nu}=\partial_\mu A^\text{pure}_\nu-\partial_\nu A^\text{pure}_\mu-ig\left[A^\text{pure}_\mu,A^\text{pure}_\nu\right]=0.
\end{equation}
It can then be written in the form
\begin{equation}
A_\mu^\text{pure}(x)=\frac{i}{g}\,U_\text{pure}(x)\partial_\mu U^{-1}_\text{pure}(x),
\end{equation}
where $U_\text{pure}(x)$ is some unitary matrix. From the gauge transformation law of this matrix 
\begin{equation}\label{Ugauge}
U_\text{pure}(x)\mapsto\tilde U_\text{pure}(x)=U(x)U_\text{pure}(x)
\end{equation}
and Eq. \eqref{Agauge}, it is easy to obtain the gauge transformation laws of the pure-gauge and physical terms
\begin{align}
A_\mu^\text{pure}(x)&\mapsto\tilde A_\mu^\text{pure}(x)=\nonumber\\
&\hspace{1cm}U(x)\left[A_\mu^\text{pure}(x)+\frac{i}{g}\,\partial_\mu\right]U^{-1}(x),\label{Apureg}\\
A_\mu^\text{phys}(x)&\mapsto\tilde A_\mu^\text{phys}(x)=U(x)A_\mu^\text{phys}(x)U^{-1}(x).\label{Aphysg}
\end{align}
In particular, note that $A^\text{phys}_\mu(x)$ transforms as an internal tensor just like any other physical fields. It is therefore natural to treat the physical term as a dynamical field, \emph{i.e.} as part of the generic set of fields $\phi(x)$, and to treat the pure-gauge term as an external field. 

Note also that rewriting the decomposition \eqref{Adecomp} in a more explicit form
\begin{align}
(-igA_\mu)^a_{\phantom{a}b}&=(U_\text{pure})^a_{\phantom{a}a'}\partial_\mu (U^{-1}_\text{pure})^{a'}_{\phantom{a}b}\nonumber\\
&\qquad+(U_\text{pure})^a_{\phantom{a}a'}(-ig\hat A^\text{phys}_\mu)^{a'}_{\phantom{a}b'} (U^{-1}_\text{pure})^{b'}_{\phantom{b}b}
\end{align}
with $\hat A^\text{phys}_\mu\equiv U^{-1}_\text{pure}A^\text{phys}_\mu U_\text{pure}$, is reminiscent of the tetrad formalism in general relativity \cite{deFelice:1990hu,Unzicker:2005in}
\begin{equation}
\Gamma^\lambda_{\mu\nu}=e^\lambda_a\partial_\mu e^a_\nu+e^\lambda_a\omega^a_{\mu b}e^b_\nu.
\end{equation}
In some sense, the fields  $-igA_\mu(x)$, $U_\text{pure}(x)$ and $-ig\hat A^\text{phys}_\mu(x)$ can be thought of as the analogs of the Christoffel symbol $\Gamma_\mu(x)$, the vierbein $e(x)$ and the spin connection $\omega_\mu(x)$, respectively.

\subsection{Gauge-covariant approach revisited}

Once more, consider for simplicity that the fields $\phi(x)$ transform as internal vectors under gauge transformations. We propose to think of the Lagrangian as a function of the gauge-covariant physical fields and their pure-gauge covariant derivatives\footnote{Note in particular that $D^\text{pure}_\mu U_\text{pure}(x)=0$.} $D^\text{pure}_\mu=\partial_\mu-igA^\text{pure}_\mu$
\begin{equation}\label{noncovariantform}
 \uL(x)=f''[\phi(x),D^\text{pure}_\mu\phi(x)].
 \end{equation}
Such a rewriting is always possible since a gauge-invariant Lagragian involves the field $A_\mu(x)$ only through covariant derivatives and field strengths
\begin{align}
D_\mu&=D^\text{pure}_\mu-igA^\text{phys}_\mu,\\
F_{\mu\nu}&=\mathcal D^\text{pure}_\mu \! A^\text{phys}_\nu-\mathcal D^\text{pure}_\nu \! A^\text{phys}_\mu-ig\left[A^\text{phys}_\mu,A^\text{phys}_\nu\right],
\end{align}
where the pure-gauge covariant derivative in the adjoint representation is given by $\mathcal D^\text{pure}_\mu=\partial_\mu-ig[A^\text{pure}_\mu,\quad]$. Note that, owing to Eq. \eqref{Apuredef}, the pure-gauge covariant derivatives commute. They are therefore the most natural gauge-invariant extensions of the ordinary partial derivatives $\partial_\mu$.

In this approach the geometrically consistent expression for the variation of the fields is obviously
\begin{equation}
\phi'(x')=\mathcal W_\text{pure}(x+\delta x,x)\left[\phi(x)+\Delta_\text{pure}\phi(x)\right],
\end{equation}
where the pure-gauge infinitesimal Wilson line and the pure-gauge covariant total variation are respectively defined as
\begin{subequations}
\begin{align}
\mathcal W_\text{pure}(x+\delta x,x)&=1+igA^\text{pure}_\mu(x)\,\delta x^\mu,\\
\Delta_\text{pure}\phi(x)&=\delta\phi(x)+D^\text{pure}_\mu\phi(x)\,\delta x^\mu.
\end{align}
\end{subequations}
The Euler-Lagrange equations and conserved Noether currents then take the form
\begin{align}
0&=\frac{\partial\uL}{\partial(D^\text{pure}_\mu\phi)}\!\stackrel{\leftarrow}{D}\!\!\!\!\!\phantom{\partial}^\text{pure}_\mu-\frac{\partial\uL}{\partial\phi},\label{ELeq}\\
\mathcal J^\mu&=\frac{\partial\uL}{\partial(D^\text{pure}_\mu\phi)}\,\Delta_\text{pure}\phi\nonumber\\
&\quad-\left[\frac{\partial\uL}{\partial(D^\text{pure}_\mu\phi)}\,D^\text{pure}_\nu\phi-\delta^\mu_\nu \uL\right]\delta x^\nu.
\end{align}

Considering infinitesimal translations of the space-time coordinates with the fields physically unchanged
\begin{equation}
\delta x^\mu=\varepsilon^\mu,\qquad\Delta_\text{pure}\phi(x)=0,
\end{equation}
we obtain from the corresponding (infinitesimal) Noether current $\mathcal J^\mu=T^{\mu\nu}\varepsilon_\nu$ the gauge-invariant canonical energy-momentum tensor
\begin{equation}\label{giEM}
T^{\mu\nu}=\frac{\partial\uL}{\partial(D^\text{pure}_\mu\phi)}\,D^\nu_\text{pure}\phi-g^{\mu\nu} \uL.
\end{equation}
Considering now infinitesimal Lorentz transformations $\Lambda^\mu_{\phantom{\mu}\nu}=\delta^\mu_\nu-\omega^\mu_{\phantom{\mu}\nu}$ with $\omega_{\mu\nu}=-\omega_{\nu\mu}$, the coordinate and field variations are given by
\begin{equation}
\delta x^\mu=\omega^\mu_{\phantom{\mu}\nu}x^\nu,\qquad\Delta_\text{pure}\phi(x)=-\frac{i}{2}\,\omega_{\alpha\beta}S^{\alpha\beta}\phi(x),
\end{equation}
where $S^{\alpha\beta}$ is the appropriate (antisymmetric) spin matrix. From the corresponding (infinitesimal) Noether current $\mathcal J^\mu=\frac{1}{2}\,M^{\mu\nu\rho}\omega_{\nu\rho}$, we obtain the gauge-invariant canonical generalized angular-momentum tensor
\begin{equation}\label{giGAM}
M^{\mu\nu\rho}=-i\,\frac{\partial\uL}{\partial(D^\text{pure}_\mu\phi)}\,S^{\nu\rho}\phi(x)+(x^\nu T^{\mu\rho}-x^\rho T^{\mu\nu}).
\end{equation}
The crucial difference with the standard treatment is simply the use of the geometrically consistent total variation $\Delta_\text{pure}\phi(x)$ instead of $\Delta\phi(x)$.

There is therefore a simple rule of thumb for reconciling the standard canonical formalism with the gauge symmetry: it suffices to replace formally $A_\mu$ by $A^\text{phys}_\mu$ and $\partial_\mu$ by the appropriate pure-gauge covariant derivative in any gauge-dependent expression.

\subsection{Application to QCD}

Using the decomposition \eqref{Adecomp}, the QCD Lagrangian can be thought of as made of three gauge-invariant terms $\uL_\text{QCD}=\uL_\text{D}+\uL_\text{YM}+\uL_\text{int}$, where the so-called Dirac, Yang-Mills and interaction terms are given by
\begin{subequations}
\begin{align}
\uL_\text{D}&=\barpsi(i\!\stackrel{\leftrightarrow}{\,\uslash \!D}\!\!\!\!\!\phantom{\partial}^\text{\,pure}-m)\psi,\\
\uL_\text{YM}&=-\frac{1}{2}\,\uTr[F^{\alpha\beta}F_{\alpha\beta}],\\
\uL_\text{int}&=g\barpsi\uslash A^\text{phys}\psi\label{Lint}.
\end{align}
\end{subequations}
We used for convenience the notation $\stackrel{\leftrightarrow}{a}\,=\frac{1}{2}\,(\stackrel{\rightarrow}{a}-\stackrel{\leftarrow}{a})$. 

From the Euler-Lagrange equations \eqref{ELeq}, we recover the standard QCD equations of motion
\begin{subequations}
\begin{align}
0&=\frac{\partial\uL}{\partial(\stackrel{\rightarrow}{D}\!\!\!\!\!\phantom{\partial}^\text{pure}_\mu\psi)}\!\stackrel{\leftarrow}{D}\!\!\!\!\!\phantom{\partial}^\text{pure}_\mu-\frac{\partial\uL}{\partial\psi}\nonumber\\
&=\barpsi (i\!\stackrel{\leftarrow}{\,\uslash\!D}\!+m),\\
0&=\stackrel{\rightarrow}{D}\!\!\!\!\!\phantom{\partial}^\text{pure}_\mu\frac{\partial\uL}{\partial(\barpsi\!\stackrel{\leftarrow}{D}\!\!\!\!\!\phantom{\partial}^\text{pure}_\mu)}-\frac{\partial\uL}{\partial\barpsi}\nonumber\\
&=(i\!\stackrel{\rightarrow}{\,\uslash\!D}\!-m)\psi,\\
0&=\mathcal D^\text{pure}_\nu\frac{\partial\uL}{\partial(\mathcal D_\nu^\text{pure}\!A^\text{phys}_\mu)}-\frac{\partial\uL}{\partial A^\text{phys}_\mu}\nonumber\\
&=2(\mathcal D_\nu F^{\nu\mu})^a_{\phantom{a}b}+g\barpsi_b\gamma^\mu\psi^a,
\end{align}
\end{subequations}
where $a,b$ are internal-space indices. For the gauge-invariant canonical energy-momentum tensor \eqref{giEM}, we obtain
\begin{equation}
T^{\mu\nu}=T^{\mu\nu}_q+T^{\mu\nu}_g-g^{\mu\nu}\uL,
\end{equation}
where the quark and gluon contributions are given by
\begin{subequations}\label{QCDEM}
\begin{align}
T^{\mu\nu}_q&=i\,\barpsi \gamma^\mu\!\!\stackrel{\leftrightarrow}{D}\!\!\!\!\!\phantom{\partial}^\nu_\text{pure}\psi,\\
T^{\mu\nu}_g&=-2\,\uTr[F^{\mu\alpha}\mathcal D^\nu_\text{pure} A^\text{phys}_\alpha].
\end{align}
\end{subequations}
Similarly, for the gauge-invariant canonical generalized angular-momentum tensor \eqref{giGAM}, we obtain
\begin{equation}
\begin{split}
M^{\mu\nu\rho}&=M^{\mu\nu\rho}_{q,\text{spin}}+M^{\mu\nu\rho}_{q,\text{OAM}}+M^{\mu\nu\rho}_{g,\text{spin}}+M^{\mu\nu\rho}_{g,\text{OAM}}\\
&\quad -x^{[\nu}g^{\rho]\mu}\uL,
\end{split}
\end{equation}
where the spin and orbital angular momentum (OAM) contributions of quarks and gluons are given by
\begin{subequations}\label{QCDGAM}
\begin{align}
M^{\mu\nu\rho}_{q,\text{spin}}&=\frac{1}{2}\,\barpsi\{\gamma^\mu,\Sigma^{\nu\rho}\}\psi=\frac{1}{2}\,\epsilon^{\mu\nu\rho\sigma}\,\barpsi\gamma_\sigma\gamma_5\psi,\\
M^{\mu\nu\rho}_{q,\text{OAM}}&=i\,\barpsi \gamma^\mu x^{[\nu}\!\!\stackrel{\leftrightarrow}{D}\!\!\!\!\!\phantom{\partial}^{\rho]}_\text{pure}\psi,\\
M^{\mu\nu\rho}_{g,\text{spin}}&=-2\,\uTr[F^{\mu[\nu} A^{\rho]}_\text{phys}],\label{Gspin}\\
M^{\mu\nu\rho}_{g,\text{OAM}}&=-2\,\uTr[F^{\mu\alpha}x^{[\nu} \mathcal D^{\rho]}_\text{pure}A^\text{phys}_\alpha],
\end{align}
\end{subequations}
with $\epsilon_{0123}=+1$ and the notation $a^{[\mu}b^{\nu]}=a^\mu b^\nu-a^\nu b^\mu$.

The expressions \eqref{QCDEM} and \eqref{QCDGAM} coincide with the gauge-invariant canonical decompositions of the proton momentum and spin proposed originally by Chen \emph{et al.} \cite{Chen:2008ag,Chen:2011zzh}, and put in a Lorentz-covariant form by Wakamatsu \cite{Wakamatsu:2010cb,Lorce:2012rr}. We therefore demonstrated here that the \emph{ad hoc} expressions of Chen \emph{et al.} can actually be derived from the canonical formalism and the Noether's theorem, once these are reconciled with the gauge symmetry. Note that they can also be derived from the standard Noether's theorem when non-standard Lorentz transformation laws for the fields are considered \cite{Guo:2012wv,Guo:2013jia}.

\subsection{Stueckelberg symmetry}

By construction, the decomposition \eqref{Adecomp} is gauge invariant, \emph{i.e.} one has $\tilde A_\mu(x)=\tilde A_\mu^\text{pure}(x)+\tilde A_\mu^\text{phys}(x)$. It is also consistent with the Lorentz symmetry, as discussed in detail in Ref.~\cite{Lorce:2012rr}. However, it is not unique since we still have some freedom in defining exactly what we mean by `pure-gauge' and `physical'. The reason is that decomposing the gauge field into two parts automatically introduces an additional local symmetry to the Lagrangian. The new symmetry has the same group structure as the gauge symmetry, but acts only on the pure-gauge and physical parts of the gauge field
\begin{subequations}\label{stueck}
\begin{align}
A^\text{pure}_\mu(x)&\mapsto A^{\text{pure},g}_\mu(x)=\nonumber\\
&\hspace{-0.75cm}A^\text{pure}_\mu(x)+\frac{i}{g}\,U_\text{pure}(x)U_0^{-1}(x)\left[\partial_\mu U_0(x)\right]U^{-1}_\text{pure}(x),\\
A^\text{phys}_\mu(x)&\mapsto A^{\text{phys},g}_\mu(x)=\nonumber\\
&\hspace{-0.75cm}A^\text{phys}_\mu(x)-\frac{i}{g}\,U_\text{pure}(x)U_0^{-1}(x)\left[\partial_\mu U_0(x)\right]U^{-1}_\text{pure}(x),
\end{align}
\end{subequations}
where $U_0(x)$ is a Stueckelberg unitary matrix. At the level of the matrices $U_\text{pure}(x)$, this transformation reads
\begin{equation}\label{Ustueck}
U_\text{pure}(x)\mapsto U^g_\text{pure}(x)=U_\text{pure}(x)U^{-1}_0(x).
\end{equation}
While the ordinary gauge transformation acts on the left of $U_\text{pure}(x)$ as in Eq.~\eqref{Ugauge}, this new transformation acts on the right. It is therefore important to distinguish them. Since the pure-gauge term $A^\text{pure}_\mu(x)$ plays a role somewhat similar to the derivative of the Stueckelberg field, we refer to this transformation as the Stueckelberg (gauge) transformation \cite{Lorce:2012rr}. This symmetry is reminiscent of the local Lorentz symmetry in the tetrad formalism of general relativity and the dual symmetry in gauge theories \cite{Chan:1997cv,Chan:2011tc,Baker:2011dg}. Note also that the Stueckelberg symmetry has no global counterpart, as one can see from Eq.~\eqref{stueck}. This means that the decomposition of the gauge field into pure-gauge and physical contributions does not introduce new conserved currents in the theory \cite{AlKuwari:1990db}. 

The Stueckelberg symmetry is a bit problematic in the sense that one can write in principle (infinitely) many Lagrangians equivalent to the original one, just by changing the explicit expressions for $A^\text{pure}_\mu(x)$ and $A^\text{phys}_\mu(x)$. To single out a particular Lagrangian in practice, one can add a gauge-invariant term that breaks the Stueckelberg symmetry. This amounts to constraining further $A^\text{phys}_\mu(x)$. One can use for example the light-front constraint $A^+_\text{phys}(x)=0$, the Coulomb constraint $\vec{\mathcal D}^\text{pure}\cdot\vec A^\text{phys}(x)=0$, the Fock-Schwinger constraint $x\cdot A^\text{phys}(x)=0$, or any other physical constraint that specifies the two physical degrees of freedom. It is important to keep in mind that, despite appearances, this procedure does not fix the gauge since $A^\text{pure}_\mu(x)$ also contributes to $A_\mu(x)$. For a more detailed discussion concerning the relation between Stueckelberg and gauge transformations, see Ref. \cite{Lorce:2013bja}. 

Explicit realizations of the decomposition \eqref{Adecomp} clarifies the physical meaning of the Stueckelberg symmetry. These realizations are essentially non-local expressions of the gauge potential $A_\mu$. The gauge symmetry is preserved in these non-local expressions thanks to compensating phase factors. In many cases, these phase factors combine into a Wilson line whose path dependence is at the origin of the Stueckelberg dependence \cite{Lorce:2012ce}. In other words, breaking explicitly the Stueckelberg symmetry amounts in many cases to determining the path of the Wilson lines. This path dependence has a physical relevance as demonstrated \emph{e.g.} by the Aharonov-Bohm effect \cite{Aharonov:1959fk} and the possibility to access the transverse canonical momentum of partons in the TMD factorization framework, see \emph{e.g.} \cite{Collins:2011zzd} and references therein. Note however that Stueckelberg dependence is more general than mere path depedence, because in certain explicit realizations, like \emph{e.g.} with the Coulomb constraint $\vec{\mathcal D}^\text{pure}\cdot\vec A^\text{phys}(x)=0$, the phase factors cannot be combined into a simple (path-dependent) Wilson line. In this sense, the approach based on the Coulomb constraint is path independent, though still Stueckelberg dependent. Note also that the Coulomb gauge is plagued by the issue of Gribov ambiguities in non-abelian gauge theories \cite{Gribov:1977wm}. On the contrary, the so-called contour gauges, which include the light-front and axial gauges, are known to be free of these Gribov ambiguities \cite{Bassetto:1983rq,Ivanov:1985np}, but suffer from other pathologies already at the perturbative level, like \emph{e.g.} the presence of divergences and/or the existence of preferred frames mirroring the effects of ghosts in covariant gauges \cite{Burnel:2008zz}.  At the non-perturbative level, these issues may have an impact similar to the Gribov copies. These remarks are naturally expected to apply to the Stueckelberg fixing procedure as well. In practice, the Chen \emph{et al.} approach should better be considered as a perturbative construction.

It is the actual physical process that determines the phase factors or the shape of the Wilson lines, and in turn which constraint on $A^\text{phys}_\mu(x)$ to use. Phase factors are necessary to preserve the gauge invariance, but favors a particular gauge constraint, the one in which they reduce to the identity or, equivalently, the one in which $A^\text{pure}_\mu(x)$ vanishes. Any gauge will of course give the same numerical answer for the physical observable, but the \emph{physical interpretation} of the latter will be the simplest in the gauge favored by the phase factor. The archetypical example is deep-inelastic scattering where the factorization theorem forces the Wilson lines entering the definition of the parton distribution functions to run along the light-front direction. At leading twist, these parton distribution functions can be interpreted as linear combinations of parton probabilities in the light-front gauge $A^+(x)=0$. The decomposition \eqref{Adecomp} simply allows one to extend this interpretation to any gauge, provided that one defines the physical term by the constraint $A^+_\text{phys}(x)=0$.

\section{Gauge-invariant canonical formalism}\label {sec4}

We present in this section an alternative to the gauge-covariant canonical formalism, which we refer to as the gauge-invariant canonical formalism. We show that these two formalisms are formally equivalent.

\subsection{Dirac variables}

Dirac soon realized that one of the main obstacles in the quantization of a gauge theory is the gauge dependence of the fields. He therefore built from the old gauge-variant fields new gauge-invariant fields that will play the role of the dynamical variables in the canonical formalism. In this spirit, we make use of the matrices $U_\text{pure}(x)$ to construct the field variables
\begin{equation}\label{generalizedDiracvariable}
\hat\phi(x)=U^{-1}_\text{pure}(x)\phi(x).
\end{equation}
For simplicity, we considered once more that the fields $\phi(x)$ transform as internal vectors under gauge transformations, but this expression can easily be adapted to any sort of internal tensors. Similarly, for the fields like $A_\mu(x)$ that do not transform as internal tensors under gauge transformations, we define
\begin{equation}\label{generalizedDiracvariable2}
\hat A_\mu(x)=U^{-1}_\text{pure}(x)\left[A_\mu(x)+\frac{i}{g}\,\partial_\mu\right]U_\text{pure}(x).
\end{equation}
From the gauge transformation law \eqref{Ugauge} of $U_\text{pure}(x)$, we see that the fields $\hat\phi(x)$ and $\hat A_\mu(x)$ are by construction gauge invariant. We refer to them as generalized Dirac variables\footnote{The original Dirac variables were constructed in the context of QED with the Coulomb constraint $\vec\nabla\cdot\vec{\hat A}(x)=0$.}.

We stress that, despite appearances, Eqs.~\eqref{generalizedDiracvariable} and~\eqref{generalizedDiracvariable2} do not represent gauge transformations. In practice, the matrices $U_\text{pure}(x)$ can be expressed in terms of the gauge field $A_\mu(x)$ \cite{Lorce:2012ce}, and can then be thought of as dressing fields. From a geometrical point of view, $U_\text{pure}(x)$ simply determines a reference configuration in the internal space. The gauge-invariant fields $\hat\phi(x)$ then represent ``physical'' deviations from this reference configuration.

\subsection{Gauge-invariant approach}

In the gauge-invariant canonical formalism, the Lagrangian is thought of as a function of the generalized Dirac variables $\hat\phi(x)$ (including from now on $\hat A_\mu(x)$ as well) and their ordinary derivatives
\begin{equation}
\uL(x)=f'''[\hat\phi(x),\partial_\mu\hat\phi(x)].
\end{equation}
Since the generalized Dirac variables are gauge invariant, we can simply apply the standard approach of Section \ref{standardapproach}. 

The rule of thumb is particularly simple: it suffices to add a hat to the fields $\phi(x)$ wherever they appear. The obtained Euler-Lagrange equations and conserved Noether currents are then automatically gauge-invariant.

\subsection{Equivalence with the gauge-covariant approach}

The gauge-covariant and invariant canonical formalisms are essentially equivalent. Indeed, noting that\footnote{For the gauge field, we have $\hat A_\mu(x)=U^{-1}_\text{pure}(x)A^\text{phys}_\mu(x)U_\text{pure}(x)$ and $\partial_\nu\hat A_\mu(x)=U^{-1}_\text{pure}(x)[\mathcal D^\text{pure}_\nu A^\text{phys}_\mu(x)]U_\text{pure}(x)$.}
\begin{equation}
\partial_\mu\hat\phi(x)=U^{-1}_\text{pure}(x)D^\text{pure}_\mu\phi(x),
\end{equation}
the Lagrangian can be rewritten in the following form
\begin{equation}
\uL(x)=f'''[U^{-1}_\text{pure}(x)\phi(x),U^{-1}_\text{pure}(x)D^\text{pure}_\mu\phi(x)].
\end{equation}
Then, thanks to the gauge symmetry of the Lagrangian, we are assured that all the matrices $U_\text{pure}(x)$ disappear in the final expression so that we can write the Lagrangian in the form of Eq.~\eqref{noncovariantform}.

In other words, one can switch between gauge-covariant and invariant canonical formalisms by a mere change of variables. Nonetheless, because of the similarity between the Schwinger approach in gauge theories and the background field method in general relativity (and other metric theories), our new gauge-covariant canonical formalism appears more suited in these contexts. This also means that the issue of uniqueness raised by the Stueckelberg symmetry affects the gauge-invariant canonical formalism as well. Indeed, we see from Eq.~\eqref{Ustueck} that, even if the generalized Dirac variables are gauge invariant, they are not Stueckelberg invariant
\begin{equation}
\hat\psi(x)\mapsto\hat\psi^g(x)=U_0(x)\hat\psi(x).
\end{equation}
There is the same freedom in defining precisely $\hat\phi(x)$ as in defining $A^\text{phys}_\mu(x)$. The existence of entire class of composite fields was already pointed out by Dirac and Steinmann \cite{Dirac:1955uv,Steinmann:1983ar,Steinmann:1985id}.

In particular, we note that the Dirac's gauge-invariant formulation of QED  is equivalent to the Chen \emph{et al.} approach, since both make use of the explicit dressing matrices $U_\text{pure}(x)=e^{ie\frac{\vec\nabla\cdot\vec A}{\vec\nabla^2}(x)}$ leading to the Stueckelberg-fixing constraint $\vec\nabla\cdot\vec A^\text{phys}(x)=0$. This particular choice makes the Coulomb gauge $\vec\nabla\cdot\vec A(x)=0$ special. In that gauge, the gauge-fixed fields coincide with the gauge-invariant ones $\phi(x)|_{\vec\nabla\cdot\vec A(x)=0}=\hat\phi(x)$. For this reason, the gauge-covariant and invariant canonical formalisms can be interpreted as gauge-invariant extensions of the standard canonical formalism.

\section{Conclusion}\label{sec5}

In this letter, we showed that separating explicitly the physical and unphysical degrees of freedom in the gauge potential allows one to reconcile in a natural way the Euler-Lagrange equations and the Noether's theorem of the standard canonical formalism with the gauge symmetry. Applying this formalism to QCD, we derived canonically the gauge-invariant operators proposed earlier by Chen \emph{et al.} in the context of the proton spin decomposition. Finally, we demonstrated the formal equivalence between our formalism and the Dirac's gauge-invariant canonical formalism. 

Because of the similarity between the approach adopted here and the background field method, we believe that the formalism developed in this letter should also be relevant to general relativity and, more generally, to any metric theories.

\section*{Acknowledgements}
In this study, I greatly benefited from numerous discussions with E. Leader, M. Wakamatsu and F. Wang on former works. I am also grateful to A. Courtoy for drawing my attention to the concept of Dirac variables. This work was supported by the P2I (``Physique des deux Infinis'') network.

\end{document}